\documentclass[10pt]{iopart}

\usepackage{graphicx}

\begin{document}
\title{Quantum Noise Locking}
\date{\today}

\author{Kirk McKenzie\dag, Eugeniy Mikhailov\ddag, Keisuke Goda\ddag, Ping Koy Lam\S, Nicolai Grosse\S, Malcolm B. Gray\dag, Nergis Mavalvala\ddag \ and David E. McClelland\dag, } 

\address{\dag\ Center for Gravitational Physics, Department of Physics, Faculty of Science, The Australian National University, ACT 0200, Australia}

\address{\ddag\ LIGO Laboratory, Massachusetts Institute of
Technology, Cambridge, 02139, USA}

\address{\S\ Quantum Optics Group, Department of Physics, Faculty of Science, The Australian National University, ACT 0200, Australia}

\begin{abstract}
Quantum optical states which have no coherent amplitude, such as squeezed vacuum states, can not rely on standard readout techniques to generate error signals for control of the quadrature phase. Here we investigate the use of asymmetry in the quadrature variances to obtain a phase-sensitive readout and to lock the phase of a squeezed vacuum state, a technique which we call noise locking (NL). We carry out a theoretical derivation of the NL error signal and the associated stability of the squeezed and anti-squeezed lock points. Experimental data for the NL technique both in the presence and absence of coherent fields are shown, including a comparison with coherent locking techniques. Finally, we use NL to enable a stable readout of the squeezed vacuum state on a homodyne detector.      
\end{abstract}
 \maketitle

\section{Introduction}

Quantum optical states are a much utilized resource required to perform quantum non-demolition readout \cite{Grangier} and to enhance the sensitivity of quantum noise limited applications \cite{Caves}. The application of quantum states requires, firstly, the generation of quality quantum states and, secondly, the readout and control of the phase of the states, such that the appropriate quadrature is measured. For the readout and control of bright squeezed states, RF modulation/demodulation or DC readout techniques can be used to stabilize the phase \cite{Chelkowski,Buchler}. For quantum states that do not have a coherent amplitude, such as squeezed vacuum states  \cite{Wu, Ou, Breitenbach} and quantum entanglement without  a coherent carrier, the usual optical phase locking techniques are not available.  In this situation the only phase-sensitive readout is the quadrature-dependent fluctuations. 


An error signal can be obtained for the case of locking the squeezed vacuum phase on a homodyne detector in an analogous way to coherent modulated techniques, for example \cite{PDH,Schnupp}. The squeezed beam phase is modulated at an RF frequency, then the noise power detected on the homodyne detector with bandwidth, $\Delta \omega$, is demodulated. This produces an error signal which has zero crossings at both the minimum and maximum variance points. This technique, which we refer to as quantum noise locking or noise locking (NL), has application in the locking of the phase of quantum states that have non-polar symmetric phase space distribution functions. This technique has been used already in squeezed vacuum experiments see, for example, \cite{Polzik,laurat1,McKenzie}. 

Moreover, in applications where the properties of squeezed vacuum states are desirable, the NL technique is the only choice for phase control\footnote{One could conceive a scheme where a 'vacuum' squeezed state is generated with coherent phase modulation sidebands without a coherent carrier amplitude imposed on it. Such an experiment has not been demonstrated to our knowledge.}. One such application maybe in gravitational wave interferometry where low frequency squeezing at the optimum phase is required. The relative immunity of squeezed vacuum states to classical noise sources can result in squeezing being produced in the audio frequency gravitational wave detection band \cite{McKenzie}. 

In this paper we investigate NL theoretically and experimentally for two different systems. The first system is the phase control of a squeezed vacuum state on a balanced homodyne detector. The second, which provides an alternative experimental demonstration of NL, is the control of the phase of two coherent beams which interfere on a balanced beamsplitter. This second system shares quadrature dependent noise with a squeezed state on a homodyne detector, since the differential phase shift of the fields changes the output power and the associated shot noise varies accordingly. This gives maximum shot noise at a bright fringe (analogous to the anti-squeezed quadrature) and minimum shot noise at the dark fringe (analogous to the squeezed quadrature).  

In Section \ref{theory}, we theoretically analyze the control of a squeezed vacuum state on a balanced homodyne detector, although the formalism is general enough to apply to other systems. We derive the NL error signal and calculate the theoretically achievable stability. Then we present theoretical results for the NL of coherent fields. We find the NL lock stability improves weakly with increasing detection bandwidth ($\Delta \omega^{1/4}$ dependence) in contrast to standard coherent locking techniques \cite{PDH}, where increasing detection bandwidth reduces the lock stability. We find that
the stability is dependent on the level of squeezing/anti-squeezing (fringe visibility in the case of coherent field NL) and also on which quadrature is locked to. Perhaps fortunately, we find the lock stability to be superior when locked to the squeezed quadrature rather than the anti-squeezed quadrature.  We also show the performance is degraded by losses and detector inefficiency, as uncorrelated vacuum fluctuations couple into the signal.

In Section \ref {experiment}, we analyze NL experimentally, using  both the coherent NL experiment setup and the squeezed vacuum on a homodyne detector. In Section \ref {C_NL}, we show results from the coherent NL experiment of the error signals and lock acquisition. We also analyze the NL stability by measuring the error signal spectra and then compare it to that of a coherent modulation locking (CML) technique. In Section \ref{SQZ_NL}, experimental results of a squeezed vacuum spectrum taken using NL for homodyne phase control are shown and we present error signal spectra which agree qualitatively with the results derived in Section \ref{theory}. We conclude  with a discussion of NL and its applications.

\section{Theory of Noise Locking}
\label{theory}
\begin{figure}[h!]
\begin{center}
\includegraphics[width=.7\columnwidth]{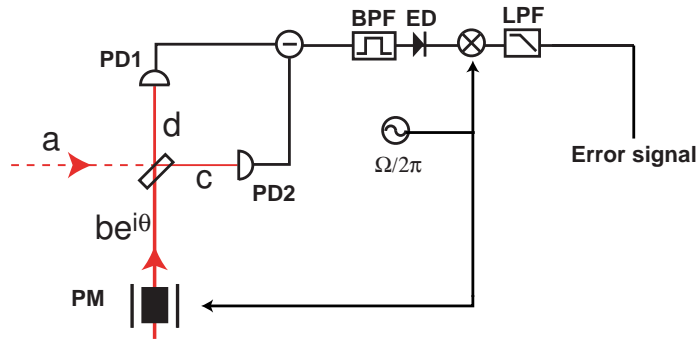}
\caption{Setup of a balanced homodyne detector with input fields, $\hat{a}(t)$ and $\hat{b}(t)$, interfering with relative phase,$\theta$, on a balanced beamsplitter.  Here the local oscillator beam $\hat{b}(t)$ passes through a phase modulator (PM) with applied sinusoidal modulation at frequency $\Omega/2\pi$. The output fields, $\hat{d}(t)$ and $\hat{c}(t)$, are incident on the photodetectors, PD1 and PD2. To derive the NL error signal, the output of the homodyne is bandpass filtered (BPF) then envelope detected (ED). The output of the envelope detector is demodulated and low pass filtered (LPF). }
\label{theory_layout}
\end{center}
\end{figure}
In this section, the error signal and lock stability of NL are derived theoretically.  Figure \ref{theory_layout} shows the input fields, $\hat{a}(t)$ and $\hat{b}(t)$,  with relative phase denoted $\theta$ that interfere on a balanced beamsplitter. These operators satisfy the standard commutation relations \cite{Scully}. The output fields, $\hat{d}(t)$ and $\hat{c}(t)$, are incident on the photodetectors; PD1 and PD2. For the two NL cases considered in this paper,  case (a); the squeezed state on the homodyne detector and case (b); the phase of two coherent fields, the input fields and photodetection differs slightly. In the case of NL of squeezed vacuum on the homodyne detector, $\hat{a}(t)$ is the squeezed state and $\hat{b}(t)$ the local oscillator (LO). The output of the balanced homodyne detector is sent to the bandpass filter (BPF). In the case of NL two coherent fields, the input fields $\hat{a}(t)$ and $\hat{b}(t)$ are set to have similar coherent amplitude. Here only one output port of the beamsplitter is detected and the output of the photodetector is sent to the BPF. All the electronics used to derive the error signal from the bandpass filter onwards are identical. The band pass filtered output is sent to an envelope detector which gives an output proportional to the real envelope of the input. This signal is then demodulated and low pass filtered to give the error signal.

\subsection{Case (a); Noise Locking squeezed vacuum on a Homodyne detector}
\subsubsection{Derivation of the noise locking error signal}
The fields in Figure \ref{theory_layout} can be decomposed into average (dc) and fluctuating (time-dependent) components, $\hat{s}(t) = \bar{s}+\delta \hat{s}(t)$ for $s = a,b,c,d$. Here average components are assumed to be real. The linearized photocurrents of PD1 and PD2 are proportional to
\begin{eqnarray}
 i_{d,c}^{\theta}(t)&= \frac{1}{2} \bigg[\bar{a}^2+\bar{b}^2\pm2\bar{a}\bar{b}\sin{\theta}+(\bar{b}\pm\bar{a}\sin{\theta})\delta X_b^{(1)} \nonumber \\
&+(\bar{a}\pm\bar{b}\sin{\theta})\delta X_a^{(1)} \pm \cos{\theta}(\bar{a}\delta X_b^{(2)}+\bar{b}\delta X_a^{(2)}) \bigg]
\end{eqnarray}
where the photocurrent $i_d^\theta(t)$ is given by the top sign and $i_c^\theta(t)$ the lower sign. The quadrature operators are defined in the standard way; $\delta X_s^{(1)} = \delta \hat{s}+\delta \hat{s}^\dagger$, $\delta X_s^{(2)} = -i(\delta \hat{s}-\delta \hat{s}^\dagger)$. The difference photocurrent, $i_{-}^\theta(t) = i_d^\theta(t)-i_c^\theta(t)$, can be decomposed into  average and fluctuating components;
\begin{eqnarray}
\bar{i}_{-}^\theta&= 2\bar{a}\bar{b}\sin{\theta} \\
\delta{i}_{-}^\theta(t) &=\bar{a} \sin{\theta}\delta X_b^{(1)}(t)+\bar{a}\cos{\theta }\delta X_b^{(2)}(t) \nonumber \\
&+\bar{b} \sin{\theta}\delta X_a^{(1)}(t)-\bar{b}\cos{\theta }\delta X_a^{(2)}(t)
\end{eqnarray}
and the variance of the difference photocurrent is given by $V_{-}^\theta =\langle(\delta i_{-}^\theta)^2\rangle$. We define 
 $V^{(1),(2)}_a,V^{(1),(2)}_b$ as the variances of the input fields $a$ and $b$, respectively. The homodyne photocurrent is bandpass filtered and envelope detected. We move into the Fourier domain, here $\tilde{V}^{(1),(2)}_{a,b}(\omega)$ represents the Fourier transform of $V^{(1),(2)}_{a,b}$. The bandwidth limited power spectrum at the output of the BPF is
\begin{eqnarray}
\tilde{S}_{-}^\theta(\omega)& =\bar{a}^2 (\tilde{V}_b^{(1)}(\omega)\sin^2{\theta} +\tilde{V}_b^{(2)}(\omega)\cos^2{\theta })\Delta \omega\nonumber \\
&+
\bar{b}^2(\tilde{V}_a^{(1)}(\omega)\sin^2{\theta}+\tilde{V}_a^{(2)}(\omega)\cos^2{\theta })\Delta \omega,
\label{v_diff}
\end{eqnarray}
where $\Delta \omega$ is the detection bandwidth. Note that we assume a perfect BPF with hard edges, experimentally this is not the case. For the condition $\bar{a}\ll\bar{b}$, the power spectrum in equation \ref{v_diff} becomes
\begin{eqnarray}
\tilde{S}_{-}^\theta (\omega)&\simeq&\bar{b}^2(\tilde{V}_a^{(1)}(\omega)\sin^2{\theta}+\tilde{V}_a^{(2)}(\omega)\cos^2{\theta})\Delta \omega
\end{eqnarray}

\begin{figure}[t]
\includegraphics[width=.8\columnwidth]{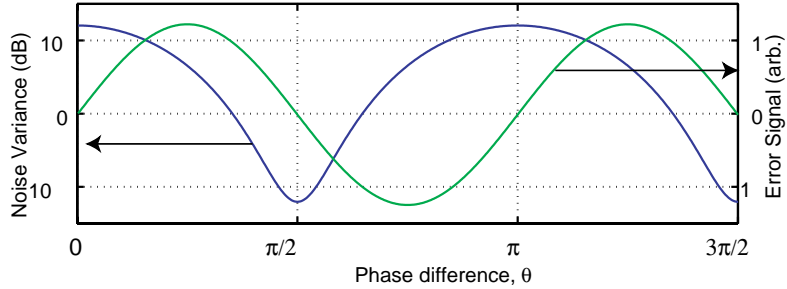}
\caption{\label{fig:fig5} The noise variance of a phase squeezed beam relative to the SNL and the NL error signal as the relative phase is varied. The error signal's zero crossing points indicate that the homodyne angle can be locked to observe both squeezing and anti-squeezing. The scale of the error signal is arbitrary. }
\end{figure}

The generation of noise locking error signals requires relative phase modulation of the input fields. We include this in the phase difference of the input fields as, $\theta = \theta_0+\theta_1\sin{\Omega t}$ where $\theta_0$ is the average phase, $\theta_1$ is the modulation depth and $\Omega$ is the modulation frequency. For small modulation depth ($\theta_1\ll\theta_0$) we make the approximation  $e^{i\theta_1\sin{\Omega t}} \simeq J_0(\theta_1)+J_1(\theta_1)e^{i\Omega t}-J_1(\theta_1)e^{-i\Omega t}$. Expanding the phase $\theta$ and neglecting the  $J_1(\theta_1)^2$ terms, we find
\begin{eqnarray}
\tilde{S}_{-}^{\theta} &\simeq \bar{b}^2\bigg[ 
\tilde{V}_a^{(1)} J_0(\theta_1)^2\sin^2\theta_0 + \tilde{V}_a^{(2)} J_0(\theta_1)^2\cos^2\theta_0 \nonumber \\&+
2J_0(\theta_1)J_1(\theta_1)\sin 2\theta_0\sin\Omega t\left(\tilde{V}_a^{(1)} - \tilde{V}_a^{(2)}\right)\bigg]\Delta \omega.
\label{Spectrum}
\end{eqnarray}
After demodulation and low-pass filtering Eqn. \ref{Spectrum} to remove the second harmonic, the NL error signal is obtained 
\begin{eqnarray}
\label{eq:errorsignal_homodyne}
\epsilon =\bar{b}^2 J_0(\theta_1)J_1(\theta_1)\sin 2\theta_0 \left(\tilde{V}_a^{(1)} - \tilde{V}_a^{(2)}\right)\Delta \omega
\end{eqnarray}
which has zero crossings at $\theta_0 =0, \pi/2$. The error signal amplitude depends on the  quadrature variances and their asymmetry. Thus if there is no asymmetry in the quadrature variances the error signal vanishes. The variance relative to the shot noise limit (SNL) and corresponding NL error signal is shown in Figure \ref{fig:fig5} for 11dB of phase squeezing. 

\subsubsection{Stability of noise locking}
\label{stability}

Since it is the variance or noise power of the detected squeezed state which is used to derive the error signal, the noise performance of the lock depends on the variance of the variance, or the \emph{noise on the noise} of the state. This can be found by taking the kurtosis, which we label $\Delta V_a^\theta$. For amplitude quadrature of the field $\hat{a}$ in the idealized squeezed state \cite{Scully}, with variance in squeezed quadrature
equal to $e^{-2R}$ and in anti-squeezed quadrature $e^{2R}$ (where $R$ is squeezing factor), the kurtosis is given by
\begin{eqnarray}
\label{eq:deltaV1}
\fl \Delta  V^{(1)}_{a} &=& \sqrt{\left\langle \left(\delta X^{(1)}_{a} - \left\langle\delta X^{(1)}_{a}\right\rangle\right)^4\right\rangle - \left\langle \left(\delta X^{(1)}_{a} - \left\langle\delta X^{(1)}_{a}\right\rangle\right)^2\right\rangle^2} = \sqrt{2}V^{(1)}_a
\end{eqnarray}
and similarly for the phase quadrature, $\Delta V_a^{(2)} = \sqrt{2}V_a^{(2)}$. Note that the kurtosis is a factor of $\sqrt{2}$ larger than the variance. 

As a measure of locking stability we express the kurtosis of the variance of the photocurrent,  $\Delta V_-^\theta$ in terms of phase fluctuations, $\Delta \theta$. We equate the kurtosis with the variance due to phase fluctuation i.e. $\Delta \tilde{V}_-^\theta=\Delta \tilde{V}_{\Delta \theta}^\theta$. Using a Taylor expansion of $\Delta \tilde{V}_{\Delta \theta}^\theta$ to second order around $\theta = \theta_0$, we find
\begin{eqnarray}
\Delta \tilde{V}^{\theta}_{-}(\theta_0)\simeq
\frac{d \tilde{V}}{d \theta}\bigg|_{\theta0} \Delta \theta +
\frac{1}{2}\frac{d^2 \tilde{V}}{d \theta^2} \bigg| _{\theta0}(\Delta \theta)^2. 
\end{eqnarray}
Expanding both sides, the equation becomes
\begin{eqnarray}
\sqrt{2}(\tilde{V}_a^{(1)} \sin^2{\theta_0} + \tilde{V}_a^{(2)} \cos^2\theta_0
)\nonumber \\ = \left|
{(\tilde{V}_a^{(1)}-\tilde{V}_a^{(2)})}\sin2\theta_0 \Delta \theta 
+      (\tilde{V}_a^{(1)}-\tilde{V}_a^{(2)}) \cos2\theta_0 (\Delta \theta)^2
\right|,
\end{eqnarray}
which when solved for  $\Delta \theta$ at the two lock points ($\theta_0 = 0, \pi/2$) gives
\begin{eqnarray}
\label{Delta_theta1}
\Delta \theta |_{\theta_0 = \pi/2} &=& \sqrt{\frac{\sqrt{2}\tilde{V}_a^{(1)}}{\tilde{V}_a^{(2)}-\tilde{V}_a^{(1)}}} \\
\label{Delta_theta2}
\Delta \theta |_{\theta_0 = 0} &=&\sqrt{\frac{\sqrt{2}\tilde{V}_a^{(2)}}{\tilde{V}_a^{(2)}-\tilde{V}_a^{(1)}}}
\end{eqnarray}

We can rewrite Eqs. (\ref{Delta_theta1}) and (\ref{Delta_theta2}) in terms of $R$, detection loss, $\lambda$ and detection bandwidth, $\Delta \omega$: With detection loss included the variance is degraded and  vacuum fluctuations are introduced, i.e. $V^{1,2}_a \rightarrow (1- \lambda) V^{1,2}_a+\lambda$.  The dependence on detection bandwidth  can be included by noting that the variance, which provides the signal for NL, is proportional to the detection bandwidth, $\Delta \omega$, and the associated noise, the kurtosis, is proportional to the square root of the detection bandwidth, $(\Delta \omega)^{1/2}$. The stability of the squeezed and anti-squeezed quadratures become 
\begin{eqnarray}
\label{Delta_theta3}
\Delta \theta |_{\theta_0 = \pi/2} &\sim& \sqrt{\frac{1+\frac{\lambda}{1-\lambda}e^{2R}}{e^{4R}-1}}\left(\frac{2}{\Delta \omega}\right)^{1/4}\\
\label{Delta_theta4}
\Delta \theta |_{\theta_0 = 0} &\sim& \sqrt{\frac{1+\frac{\lambda}{1-\lambda}e^{-2R}}{1-e^{-4R}}}\left(\frac{2}{\Delta \omega}\right)^{1/4}
\end{eqnarray}
where we have taken the case of amplitude quadrature squeezing.
\begin{figure}
\includegraphics[width=.7\columnwidth,angle=-90]{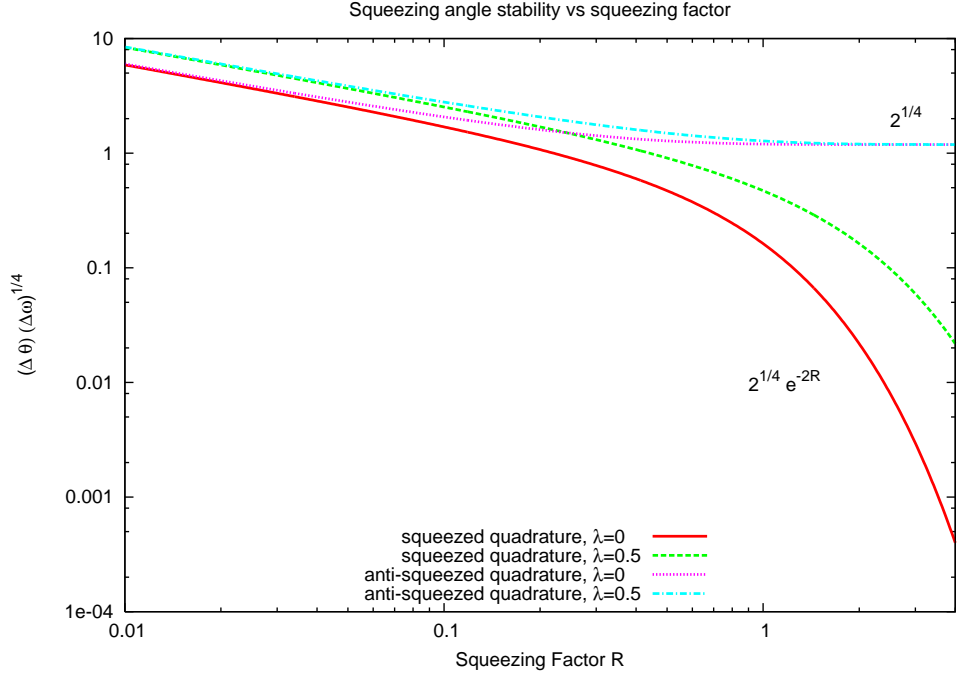}
    \caption{
	Squeezing angle stability vs squeezing factor, $R$, for the cases of
	NL to the squeezed and anti-squeezed quadrature, for
	various level of detection loss, $\lambda$.  
	\label{squeezing_stability_vs_R}
    }
\end{figure}

The stability of the two lock points are plotted as a function of squeezing factor in Figure \ref{squeezing_stability_vs_R}. For both the squeezed and anti-squeezed quadratures the stability of the lock improves as the squeezing factor is increased. This is not surprising since it is the from quadrature asymmetry that the error signal is derived. However, the stability of the two lock points, the squeezed or anti-squeezed  quadrature behave quite differently. The squeezed quadrature lock stability is perfect in the limit of perfect squeezing since the noise in variance becomes infinitely small. Note that losses and detector inefficiency mean that this will never be produced experimentally as shown with the traces for two different detector losses. The stability of the anti-squeezed quadrature lock point approaches 2$^{1/4}$ at high squeezing factor and is always higher than the squeezed quadrature lock point. Eqs. (\ref{Delta_theta3}) and  (\ref{Delta_theta4}) show that the lock stability for both quadratures improves as the detection bandwidth is increased, albeit with a weak dependence.

\subsection{Case (b); Noise Locking the phase of Coherent Fields}

The setup of NL of two coherent fields differs sightly from that of the squeezed vacuum on a homodyne detector. We set the amplitudes of the two coherent input fields to be similar in Figure \ref{theory_layout} and need detect only one output of the beamsplitter\footnote{ The phase difference between squeezed and anti-squeezed quadratures is $\pi/2$, whereas bright and dark fringes in the coherent interference experiment are separated by a phase shift of $\pi$.}. 
Following the procedure set out above, the error signal for the coherent NL is found to be
\begin{eqnarray}
\label{eq:errorsignal_c_nl}
\epsilon =\bar{a}\bar{b} J_1(\theta_1)\cos\theta_0 \Delta \omega
\end{eqnarray}
 
For comparison with the case of NL squeezed vacuum state to a homodyne detector the stability of locking to the dark ($\theta =3\pi/2$) and bright ($\theta =\pi/2$) fringes are 
\begin{eqnarray}
\label{D_theta3pi_2}
\Delta \theta |_{\theta = \frac{3\pi}{2} }  \sim \sqrt{\frac{\sqrt{2}(a-b)^2 }{ab} } 
  \left(\frac{1}{\Delta \omega}\right)^{1/4}\\
\label{D_thetapi_2}
\Delta \theta |_{\theta = \frac{\pi}{2}} \sim \sqrt{\frac{\sqrt{2}(a+b)^2 }{ab} } 
  \left(\frac{1}{\Delta \omega}\right)^{1/4}
\end{eqnarray}
Note that kurtosis in this case has the same dependence on variance as in case of squeezed state.  The functional form is the same for the dark fringe stability and the squeezed quadrature stability, Eqs. (\ref{D_theta3pi_2}) and (\ref{Delta_theta3}), respectively. Similarly for the bright fringe and anti-squeezed quadrature stabilities, Eqs. (\ref{D_thetapi_2}) and (\ref{Delta_theta4}) respectively. 
The dependence on detection bandwidth is found to be identical in the two cases.  

\section{Experimental Demonstration of Quantum Noise Locking}
\label{experiment}

An important feature of quantum noise locking is that this technique can be used in the presence of coherent fields, but, more importantly, also when no coherent amplitude is present. In this section, we show experimental demonstrations of noise locking in these two cases: (i) when coherent fields are available, as at the output of a Mach-Zehnder interferometer; and (ii) the output of a vacuum-seeded OPO, where no coherent amplitude is present.
 
\subsection{Experimental Analysis of Noise Locking of Coherent fields}
\label{C_NL}
\begin{figure}[h!tbp]
\begin{center}
\includegraphics[width=\columnwidth]{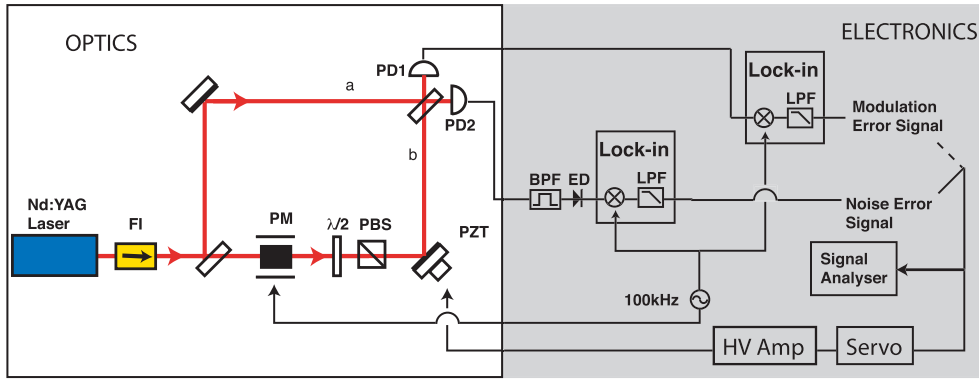}
\caption{Experimental schematic of the setup used to analyze noise locking. In the optics section; FI - Faraday isolator, PM - Phase modulator $\lambda$/2 - half-wave plate, PBS - Polarizing beam splitter, PZT - peizo electric transducer boded onto a mirror, PD1,PD2 - photo-detectors 1\&2.   In the electronics section; Lock-in - Lock-in amplifier, BPF - Band pass filter, ED - envelope detector, LPF - Low pass filter, HV amp - High voltage amplifier. }
\label{exp}
\end{center}
\end{figure}

 A schematic of the experiment is shown in Figure \ref{exp}, broken into two sections; optics and electronics. In the optical part of the experiment, 2mW from a Nd:YAG laser operating at 1064nm was split into two beams for the input fields to the second beamsplitter of the Mach-Zehnder configuration. In Figure \ref{exp}, the upper arm field corresponds to the field $\hat{a}$ whilst the lower arm corresponds to the field $\hat{b}$, since this arm contained a phase modulator (modulated at $\Omega/2\pi$=100kHz with modulation depth, $\beta \approx 0.045$~radians). This modulation was used in the derivation of the NL error signal. A mirror mounted on a piezoelectric transducer (PZT) was the actuator used in the feedback loop to control the relative phase of the input beams. A variable attenuator consisting of a $\lambda/2$ plate and a polarizing beam-splitter (PBS) was placed in the lower arm to allow the fringe visibility to be changed - to mimic changing the level of squeezing and anti-squeezing. The fringe visibility was set to 0.6 to have 6dB noise power variation on the fringe.  Both output ports of the beamsplitter were detected on matched photodetectors (PD1 and PD2) with ETX500 photodiodes, but only one photodetector (PD2) was used to derive the NL error signal. An error signal using a standard coherent modulation locking (CML) technique  was derived from photodetector (PD1). This was used for comparison with the NL technique. 

The NL error signal was produced as follows: The output of PD2 was bandpass filtered, with low frequency cutoff, $f_l$ = 2MHz and high frequency cutoff $f_h$ = 20MHz giving a detection bandwidth $\Delta \omega/2\pi= 18$MHz. The low frequency corner was designed to cut out any component of the coherent modulation signal at 100kHz by employing a low frequency corner with f$^3$ roll up. Over this frequency range most of the spectrum (5MHz - 20MHz) is shot noise limited, however below 5MHz there is some classical intensity noise present. The BPF output was then sent to an envelope detector, which had a series of amplifying stages then a diode stage, giving an output proportional to real envelope of the signal below the cut-off frequency, which in our case was 200kHz. The output of the envelope detector was then demodulated using an low frequency lock-in amplifier (Stanford Research Systems (SRS)-SR830) to give the NL error signal. The error signal was then low pass filtered to remove the second harmonic then passed through the servo and amplified before being fed back to the PZT actuator.

For the derivation of the CML error signal the photodetector output of PD1 was sent to an identical lock-in amplifier where it was demodulated then low pass filtered.  This error signal was used for comparison of NL.

\begin{figure}[h!]
\begin{center}
\includegraphics[width=\columnwidth]{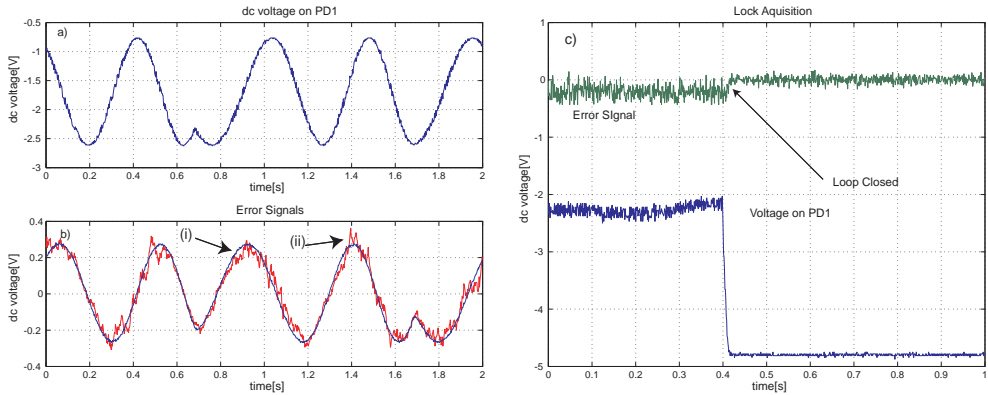} \caption{a) The DC voltage on PD1 as the fringe is scanned (the photodetector is negatively coupled, thus lower voltage corresponds to higher incident power).  b) The corresponding CML error signal (trace (i)) and the NL error signal trace (ii). c) Lock acquisition trace showing the noise locking error signal (top trace) and voltage on PD1(bottom trace). Initially, the differential phase is not controlled, about 0.4 seconds into the trace the control loop is closed and the interferometer is locked to a bright fringe on PD1 (dark fringe on PD2). Lock-in amplifier settings; LPF Time constant = 10ms, 6dB/octave for the both CML techniques. In trace a) the DC voltage has been split due the impedance of the lock-in amplifier.}
\label{esig_scan}
\end{center}
\end{figure}

Figure \ref{esig_scan} shows a) the detected optical power on PD1 (the photodetector is negatively coupled) and b) the error signals of (i) the modulation technique  and (ii) the NL error signal as the fringe was scanned. Note that the demodulation phase of the two techniques has 180$^0$ difference to give the error signals the same sign in the figure, given that the signals are derived from different beamsplitter ports. It can be seen that both error signals have zero crossing points at the bright and dark fringes. The noise of the NL error signal is much greater than that of the CML technique, which is not visible on this scale. Also, the noise on the NL error signal varies significantly over the fringe. The noise on the NL error signal is minimized at the dark fringe for PD2 (bright fringe for PD1) and maximum at the bright fringe for PD2 (dark fringe for PD1). This result agrees with the findings of Section \ref{stability} and can be seen in Figure \ref{squeezing_stability_vs_R}.

Lock acquisition using NL is shown in Figure \ref{esig_scan}  c).  The bottom  trace is the optical power on PD1 and the top trace is the NL error signal. Initially, the control loop is open. At ~0.4 seconds into the trace the control loop is closed. Here the NL error signal is quickly zeroed and the fringe on PD1 reaches the maximum value. The NL system could maintain lock indefinitely.

\begin{figure}[h!t]
\begin{center}
\includegraphics[width=0.6\columnwidth]{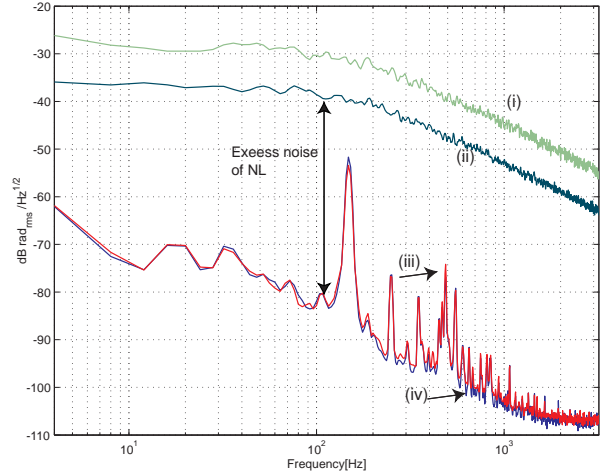}
\caption{Spectral density of the error signals whilst the interferometer is locked using CML. Traces (i) and (ii) are the NL error signal spectra of the bright and dark fringes (on PD2). Traces (iii) and (iv) are the CML error signal spectra of the bright and dark fringes (on PD1). The excess noise of the NL readout can be seen from the different amplitudes of the CML and NL error signal spectra. }
\label{nl_ml_fft}
\end{center}
\end{figure}

The noise performance of the NL was compared to the CML technique. This comparison can be seen in the spectral density of the error signals recorded on a signal analyzer (SRS-SR785), shown in Figure \ref{nl_ml_fft} whilst the system was locked using CML. The CML error signal spectral densities are shown with little difference in the spectra of the signals for the bright fringe locking [trace (iii)] and dark fringe locking [trace (iv)]. Many mechanical noise sources in the interferometer couple into the readout and can be seen in the structure shown in the error signal spectra. The NL error signal spectra are for the bright fringe on PD2  [trace (i)]; and for the dark fringe on PD2 [trace (ii)]. These NL spectra are have excess noise of greater than 40dB and show little resemblance to the CML error signal spectra.  This indicates the NL technique has substantial excess noise which buries the interferometer noise. The noise of the bright fringe NL readout is approximately 6dB larger than the dark fringe NL readout, a result which agrees qualitatively with the theoretical stability results in Section \ref{stability}.

\subsection{Application of noise locking to a squeezed vacuum state}
\label{SQZ_NL}

\begin{figure}[t!]
\begin{center}
\includegraphics[width=\columnwidth]{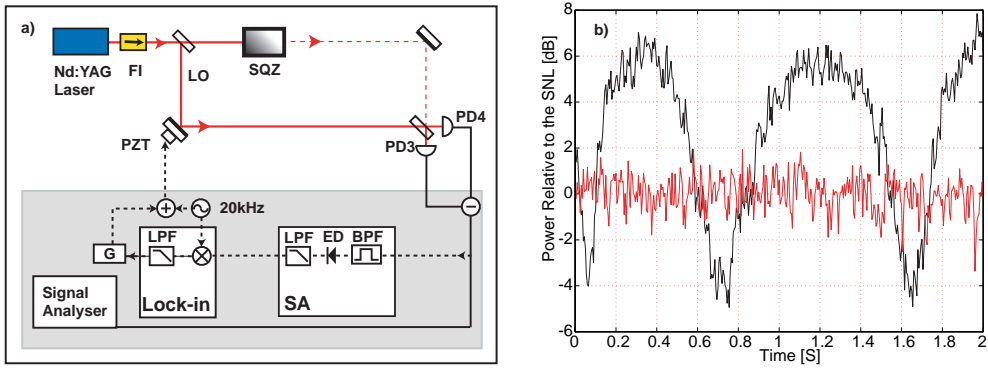}
\end{center}
\caption{a) Schematic of the squeezed vacuum experiment. b) The squeezed state at 11.2kHz as the phase of the homodyne is
varied.  RBW=1kHz, VBW=30Hz.  Electronic noise (9dB below SNL) was subtracted from
the data.} \label{OPO}
\end{figure}

Figure \ref{OPO} a) shows a schematic of the squeezed vacuum locking experiment. A Nd:YAG laser was used for a LO and to  pump a squeezed vacuum generator. Figure \ref{OPO} b) shows the noise power relative to the shot noise limit (SNL) of a squeezed vacuum beam as the LO phase is varied. The squeezing amplitude measured (after losses) is 3.6dB, which corresponds to squeeze factor R = 0.41.  

The squeezed vacuum experiment consisted of second harmonic generator used to pump an optical parametric oscillator (OPO) operating below threshold, where the squeezed vacuum state was generated.  Noise locking error signal used to control the homodyne phase was produced similarly to the technique in the experiment in the previous section. The relative phase of the homodyne detector was modulated at $\Omega/2\pi= 20$kHz using a mirror mounted on a PZT. This PZT mounted mirror was also used as the actuator in the control loop. We used a spectrum analyzer as the envelope detector (Agilent-E4407B)  at 2 MHz, with detection bandwidth (resolution bandwidth (RBW))~300 kHz and low frequency cutoff  (video bandwidth (VBW)) of  30kHz. Further experimental details can be found in \cite{McKenzie}. 

\begin{figure}[t!]
\begin{center}
\includegraphics[width=.6\columnwidth]{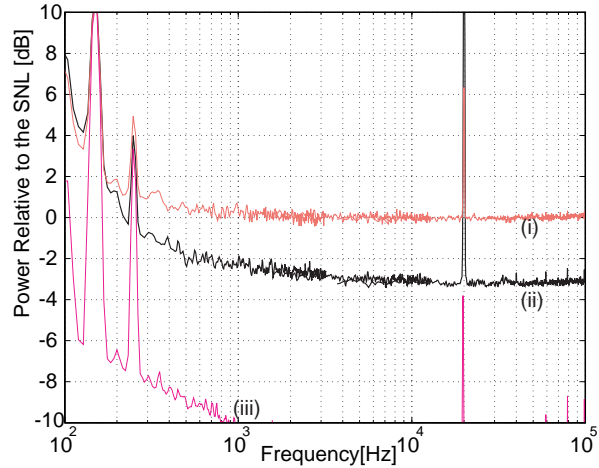}
\end{center}
\caption{ Measured noise spectra for (i) the quantum noise limit of the homodyne detection system, (ii)
the squeezed vacuum and (iii) the electronic noise of the homodyne
detection system. The electronic noise was -12dB below the quantum
noise from 10kHz-100kHz.  The 20kHz peak arises from the NL modulation.  Peaks at 50Hz harmonics are due to electrical mains supply.} \label{OPO_2}
\end{figure}

The spectra in Figure \ref{OPO_2} show squeezing from 280Hz to 100kHz [trace (ii)]; below the quantum noise limit of the homodyne detector  [trace (i)]; and the electronics noise floor [trace (iii)]. The squeezed vacuum spectra is over 3dB below the SNL over most of the frequency range but reduces at lower frequency. 
Whilst this data was taken the OPO cavity was not length controlled, since there was no coherent field in the OPO cavity that could be used as a phase reference\footnote{The cavity could have also been NL, however for the purpose of taking the data this was unnecessary.}. This meant the cavity was brought onto resonance manually and would then slowly de-tune. As the OPO cavity de-tunes the squeezed quadrature of the squeezed vacuum output changes. Also, the pump phase was not controlled, which again means the squeezed quadrature was free to evolve. Both of these issues are nulled as NL locks the LO phase to the squeezed quadrature. This property enabled the data traces to be recorded and the stability of the system was limited by the unlocked OPO, not the NL loop. 

\begin{figure}[t!]
\begin{center}
\includegraphics[width=.6\columnwidth,angle=-90]{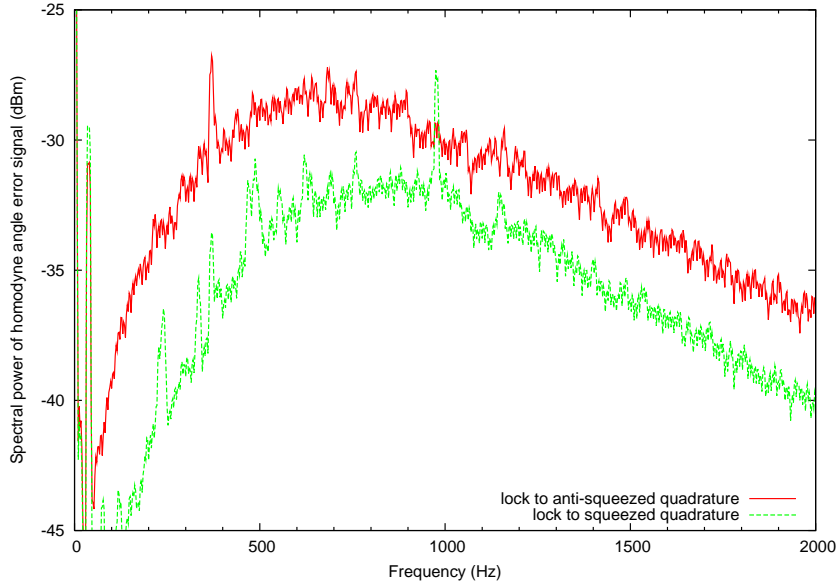}
\end{center}
\caption{The error signal spectra taken whilst a squeezed vacuum state was NL on a homodyne detector. The top trace was recorded whilst the homodyne was locked to the anti-squeezed quadrature. The lower trace was recorded whilst the homodyne was locked to the squeezed quadrature. Experimental parameters: The modulation frequency $\Omega/2\pi$ = 19.7kHz. BPF low frequency corner at $f_l$=1MHZ and high frequency corner, $f_h$=30MHz. The cutoff frequency at the output of the ED was 100kHz. The Lock-in amp had: time constant 100$\mu$S, slope 12dB/ octave} \label{esig_spectra}
\end{figure}

The NL error signal spectrum taken in-loop (whilst the squeezed state homodyne was locked using NL with a bandwidth of about 400 Hz) is shown in Figure \ref{esig_spectra}. This data was taken on a different squeezing experiment and which had slightly different experimental parameters \footnote{This data was recorded in the MIT laboratory. All other experimental data was recorded in the ANU laboratory, hence the different experimental parameters.}. The top trace was taken with the homodyne phase locked to the anti-squeezed quadrature (5dB above the SNL at 4MHz) and the bottom trace is taken whilst the the homodyne phase was locked to the squeezed quadrature (1dB below the SNL at 4MHz). The roll up of the spectrum below 1kHz is due to the loop gain suppression of the NL error point. Above the unity gain frequency of the control loop, approximately 400Hz, the spectral power of the error signal rolls off with the control loop low pass filter.
The difference in the two error signal spectral amplitude for the squeezed and anti-squeezed lock points is due to the excess noise introduced in the anti-squeezed lock. This agrees qualitatively with the result derived in Section \ref{stability} and with the experimental data for coherent fields plotted in Figure \ref{nl_ml_fft}. 

\section{Conclusions}

In this paper we have analyzed the NL technique both theoretically and experimentally. In the theory section, we derived the error signal for locking a squeezed vacuum state on a homodyne detector. The stability of NL was analyzed and it was found the stability improves with squeezing amplitude and detection bandwidth. Also, the squeezed quadrature lock stability is always superior to the anti-squeezed quadrature lock stability. We found that detector inefficiencies and losses degrade the stability of NL, since uncorrelated vacuum fluctuations are coupled into the signal. 

As expected, in the experimental analysis of NL we found that the stability of noise locking is significantly less than what can be achieved with coherent modulation locking. However, in the absence of coherent fields, the noise locking technique remains a good candidate for extracting error signals to control quadrature phases. With moderate detection bandwidth ($\Delta f = 18MHz$) and fringe visibility of 0.6, (equivalent squeezing factor of R = 0.35) the stability of NL was on the order of  40dB worse than CML technique. To improve the stability of NL two options are apparent. The first is to increase the detection bandwidth and the second is to increase the squeezing amplitude. However, both of these may be difficult experimentally. The bandwidth dependence of the stability is weak, $({1}/{\Delta \omega})^{1/4}$, as argued in Section \ref{stability}. Given most squeezing experiments have optically limited bandwidth of order 10MHz \cite{Buchler2} this will not generally be an option. Stability can also be improved by increasing the squeezing factor, however given the experimentally challenges of increasing squeezing factor (the best measured is 7dB \cite{Lam99}) this avenue is unlikely to make a significant improvement. In conclusion, though the stability coherent modulation locking is always superior to that of noise locking, noise locking can be quite an effective technique for locking of squeezed vacuum states using homodyne detection.

\section{Acknowledgments}

The authors would like to thank Julien Laurat, Nicolas Treps, Stan Whitcomb, Thomas Corbitt and Warwick Bowen for useful discussions. This research was supported in part by the Australian Research Council. We gratefully acknowledge support from National Science Foundation grants PHY-0107417 and PHY-0300345.
\\

\end{document}